\documentclass[%
aip,
amsmath,amssymb,
reprint,%
]{revtex4-1}

\usepackage{graphicx}% Include figure files
\usepackage{dcolumn}% Align table columns on decimal point
\usepackage{bm}% bold math
\usepackage[utf8]{inputenc}
\usepackage[T1]{fontenc}
\usepackage{mathptmx}
\usepackage{etoolbox}

\makeatletter
\def\@email#1#2{%
	\endgroup
	\patchcmd{\titleblock@produce}
	{\frontmatter@RRAPformat}
	{\frontmatter@RRAPformat{\produce@RRAP{*#1\href{mailto:#2}{#2}}}\frontmatter@RRAPformat}
	{}{}
}%
\makeatother

\usepackage[dvipsnames]{xcolor}
\def\mycolor{black}
\def\ct{\color{\mycolor}}
\usepackage{glossaries}
\newacronym{da}{DA}{data assimilation}
\newacronym{enkf}{EnKF}{ensemble Kalman filter}
\newacronym{enks}{EnKS}{ensemble Kalman smoother}
\newacronym{pde}{PDE}{partial differential equation}
\newacronym{rvm}{RVM}{relevance vector machine}
\newacronym{ml}{ML}{machine learning}
\newacronym{dnn}{DNN}{deep neural network}
\newacronym{ks}{KS}{Kuramoto--Sivashinsky}
\newacronym{etdrk4}{ETDRK4}{exponential time differencing fourth-order Runge-Kutta}
\newacronym{gcm}{GCM}{global climate model}
\newacronym{nwp}{NWP}{numerical weather prediction}
\newacronym{mooam}{MAOOAM}{modular arbitrary--order--ocean--atmosphere model}
\newacronym{qg}{QG}{quasi--geostrophic}
\newcommand{\doc}{manuscript}
 
\newcommand{\apriori}{\textit{a priori}}

\newcommand{\figref}[1]{Fig.~\ref{#1}}
\newcommand{\Modelerror}{Model error}
\newcommand{\modelerror}{model error}
\newcommand{\modelerrors}{model errors}%{{\color{blue}model error}}
\newcommand{\noisy}{noisy}%{{\color{red}noisy}}
\newcommand{\Noisy}{Noisy}
\newcommand{\method}{MEDIDA}
\newcommand{\stateparnull}{\ensuremath{ u }} % state parameter - continuous
\newcommand{\stateparc}{\ensuremath{ \stateparnull }} % state parameter - continuous
\newcommand{\statepar}{\ensuremath{ \bm{\stateparc} }}
\newcommand{\modelstateparc}{\ensuremath{ \stateparnull(t) }} % state parameter - continuous

\newcommand{\statedis}[1]{\ensuremath{ \bm{\stateparnull} }} % 
\newcommand{\stateobs}[1]{\ensuremath{ \statedis{}^o_{#1} }} % 
\newcommand{\statemodel}[1]{\ensuremath{ \statedis{}^m_{#1} }} % 
\newcommand{\stateanalysis}[1]{\ensuremath{ \statedis{}^a_{#1} }} % 
\newcommand{\grid}{\ensuremath{ \bm{x} }} % grid
\newcommand{\modelstateparobsc}{\ensuremath{ \stateparnull^o(t) }} % observed state parameter - continuous
\newcommand{\modelstateparmod}[1]{\ensuremath{ \bm{\stateparnull}^m( #1) }} % model state parameter - discrete

\newcommand{\modelstateparobs}[1]{\ensuremath{ \bm{\stateparnull}^o(#1) }} % model state parameter - discrete

\newcommand{\timepre}{\ensuremath{ t_{i}-\Delta t }} % Discrete time counter
\newcommand{\timepreti}{\ensuremath{ t_{i}-\Delta t_{i} }} % Discrete time counter
\newcommand{\timenow}{\ensuremath{ t_{i} }} % Discrete time counter
\newcommand{\minimize}{\ensuremath{ \text{min} }}

\newcommand{\realsetnull}{\ensuremath{ \mathbb{R} } } % Set of real 1D
\newcommand{\realsetO}[1]{\ensuremath{ \realsetnull^{#1} } } % Set of real 1D
\newcommand{\realset}[2]{\ensuremath{ \realsetO{#1 \times #2} }} % Set of real, 
\newcommand{\realsetT}[2]{\ensuremath{ \realsetO{#1 \times #2} }}% Set of real 
\newcommand{\Norm}[1]{\ensuremath{  \left\| #1 \right\| }}

\newcommand{\NormT}[1]{\ensuremath{  \Norm{#1}_2 }}

\newcommand{\inv}[1]{\ensuremath{  #1^{-1} }}

\newcommand{\transpose}[1]{\ensuremath{ {#1}^{\top} }}
\newcommand{\transposeParan}[1]{\ensuremath{ \transpose{\left(#1\right)} }}
\newcommand{\error}{\ensuremath{ \varepsilon }}

\newcommand{\timestep}{\ensuremath{\Delta t}}

\newcommand{\disNormal}[2]{\ensuremath{ \mathcal{N}\left(#1, #2\right) }}

\newcommand{\numgrid}{\ensuremath{M}}
\newcommand{\numens}{\ensuremath{N}}

\newcommand{\numsamples}{\ensuremath{n}}
\newcommand{\noiseratio}{\ensuremath{ \sigmaobs/\sigma_{u} }}
\newcommand{\sigmaobs}{\ensuremath{ \sigma_{obs} }}
\newcommand{\sigmainflation}{\ensuremath{ \sigma_{b} }}

\newcommand{\coeffsymbole}{\ensuremath{{c}}}
\newcommand{\coeff}{\ensuremath{\bm{\coeffsymbole}}}

\newcommand{\coefftrue}{\ensuremath{ {\coeff}_{s} }}
\newcommand{\coeffdiscovered}{\ensuremath{ \coeff^* }}
\newcommand{\coeffmodel}{\ensuremath{ \coeff_m }}
\newcommand{\bases}{\ensuremath{ \bm{\Phi} }}
\newcommand{\basis}{\ensuremath{ {\phi} }}
\newcommand{\basesof}[1]{\ensuremath{ \bm{\Phi}\left(#1\right) }}

\newcommand{\lincombxuObs}{\ensuremath{ \basesof{\stateobs{} } \coeff }}

\newcommand{\funmodelnull}{\ensuremath{ f }} 
\newcommand{\funtruthnull}{\ensuremath{ g }} 
\newcommand{\funerrornull}{\ensuremath{ h }}

\newcommand{\funmodel}[1]{\ensuremath{ \funmodelnull\left(#1\right) }} 
\newcommand{\funtruth}[1]{\ensuremath{ \funtruthnull\left(#1\right) }} 
\newcommand{\funerror}[1]{\ensuremath{ \funerrornull\left(#1\right) }} 

\newcommand{\funmodeldis}[1]{\ensuremath{ \bm{\funmodelnull}\left(#1\right) }} 
 
\newcommand{\funerrordis}[1]{\ensuremath{ \bm{\funerrornull}\left(#1\right) }} 

\newcommand{\errorcoeffmodel}{\ensuremath{ \error_m }}
\newcommand{\errorcoeffdiscovered}{\ensuremath{ \error^*} }

\newcommand{\numpoly}{\ensuremath{p}}
\newcommand{\numderv}{\ensuremath{d}}

\newcommand{\powpoly}{{r}}
\newcommand{\powderv}{{s}}

\newcommand{\numlibcol}{\ensuremath{q}}
\usepackage[left]{lineno}
\usepackage{tabularx}
\usepackage[caption=false]{subfig}% at some point for APS
\def\myscaletwo{0.85}%{0.95}
\def\myscalethree{1.40}

\usepackage[autolanguage]{numprint}
\nprounddigits{2}

\usepackage{hyperref}
\def\mydepository{\url{https://github.com/envfluids/MEDIDA}}

\begin{document}
	
	\preprint{AIP/123-QED}

	\title[Discovering interpretable model errors]{
		{Discovery of  interpretable} structural model errors\\
		by combining Bayesian sparse regression and data assimilation:\\ A chaotic Kuramoto-Sivashinsky test case
	}

	\author{Rambod Mojgani}
	\email{rm99@rice.edu}
	\author{Ashesh Chattopadhyay}
	%\email{akc6@rice.edu}
	\author{Pedram Hassanzadeh}
	%\email{pedram@rice.edu}
	\affiliation{%
		Rice University, Houston, TX 77005, USA
	}%

	\date{\today}% It is always \today, today,
	
	\begin{abstract}
	Models of many engineering and natural systems are imperfect. The discrepancy between the mathematical representations of a true physical system and its imperfect model is called the model error. These model errors can lead to substantial differences between the numerical solutions of the model and the {state} of the system, particularly in those involving nonlinear, multi--scale phenomena. Thus, there is {increasing interest in reducing model errors, particularly {by leveraging the rapidly growing observational data to understand their physics and sources}}. Here, we introduce a framework named MEDIDA: Model Error Discovery with Interpretability and Data Assimilation. MEDIDA only requires a working numerical solver of the model and a small number of noise--free or noisy sporadic observations of the system. In MEDIDA, first the model error is estimated from differences between the observed states and model--predicted states (the latter are obtained from a number of one--time--step numerical integrations from the previous observed states). If observations are noisy, a \gls{da} technique such as \gls{enkf} {is employed to provide the analysis state} of the system, which is then used {to estimate} the model error. Finally, an equation-discovery technique, here the \gls{rvm}, a sparsity--promoting Bayesian method, is used to identify an interpretable, parsimonious, and closed--form representation of the model error. Using the chaotic \gls{ks} system as the test case, we demonstrate the excellent performance of \method\ in discovering different types of structural/parametric model errors, representing different types of missing physics, using noise--free and noisy observations.
	\end{abstract}
	
	\maketitle
	
	\glsresetall
	\begin{quotation}
	% copied from introduction.tex
	% copied from introduction.tex
	The discovery of the governing equations of physical systems is often based on the first principles, which has been the origin of most advances in science and engineering. However, in many important applications, some of the underlying {physical processes are not well understood, and therefore, are missing or poorly represented in the mathematical models of {such} systems. 
	Accordingly, these {\textit{imperfect models}} (``models'' hereafter) can merely closely track the dynamics of the {\textit{true physical system}} (``system'' hereafter), while failing to exactly represent it. \Glspl{gcm} and \gls{nwp} models are prominent examples of such models, often suffering from parametric and structural errors. The framework proposed here integrates \gls{ml} and \gls{da} techniques to discover close--form, interpretable representations of these model errors. This framework can leverage the ever-growing abundance of observational data to reduce the errors in models of nonlinear dynamical systems, such as the climate system.}
%	\Gls{da} techniques are used to fuse observation into the models. However, \gls{da} does not provide any additional information on the underlying mechanisms leading to the deviation of the model from the truth.
%	We propose an approach to leverage \gls{ml} and \gls{da} to uncover closed--form and interpretable mechanisms leading to the mismatch of observations and the corresponding models. The proposed method can further be applied to parameterization of unresolved quantities in climate and weather models.

\end{quotation}
	
%\begin{keyword}%Use showkeys class option if keyword display desired
%	\modelerror \sep
%	equation discovery \sep
%	machine learning  \sep
%	data assimilation  \sep 
%	Bayesian  \sep
%	ensemble Kalman filter 
%\end{keyword}

%\maketitle
%\glsresetall
%\end{frontmatter}

%\linenumbers

%\input{cases_table.tex}
\newcommand{\mycaseone}{\ensuremath{ \partial^2_{x}u + \partial^4_{x}u }}
\newcommand{\mycasetwo}{\ensuremath{ u\partial_{x}u + \partial^4_{x}u }}
\newcommand{\mycasethree}{\ensuremath{ u\partial_{x}u + \partial^2_{x}u }}
\newcommand{\mycasefour}{\ensuremath{ u\partial_{x}u }}
\newcommand{\mycasefive}{\ensuremath{ \partial^2_{x}u }}
\newcommand{\mycasesix}{\ensuremath{  u_{xxxx} }}
\newcommand{\mycaseseven}{\ensuremath{ 0.5 u\partial_{x}u + \partial^2_{x}u + \partial^4_{x}u }}
\newcommand{\mycaseeight}{\ensuremath{ u\partial_{x}u + 0.5 \partial^2_{x}u + \partial^4_{x}u }}
\newcommand{\mycasenine}{\ensuremath{ u\partial_{x}u + \partial^2_{x}u + 0.5 \partial^4_{x}u }}
\newcommand{\mycaseten}{\ensuremath{  2 u\partial_{x}u + 2 \partial^2_{x}u + 2 \partial^4_{x}u }}
\newcommand{\mycaseeleven}{\ensuremath{ 0.5 u\partial_{x}u + 2 \partial^4_{x}u  }}
\newcommand{\mycasetwelve}{\ensuremath{ 0.5 u\partial_{x}u + 2 \partial^4_{x}u  }}

% =====================================================
\newcommand{\mycasetwentyone}{\ensuremath{ u\partial_{x}u  +  \partial^2_{x}u +  0.1 \partial^3_{x}u + \partial^4_{x}u  }}
\newcommand{\mycasetwentytwo}{\ensuremath{ u\partial_{x}u  +  \partial^2_{x}u +  0.5 \partial^3_{x}u + \partial^4_{x}u  }}
\newcommand{\mycasetwentythree}{\ensuremath{ u\partial_{x}u  +  \partial^2_{x}u +  1.0 \partial^3_{x}u + \partial^4_{x}u  }}

\newcommand{\mycasettwentyTone}{\ensuremath{  \partial^2_{x}u  +  0.5 \partial^3_{x}u  + \partial^4_{x}u   }}
\newcommand{\mycasetwentyTtwo}{\ensuremath{ u\partial_{x}u  +  0.5 \partial^3_{x}u  + \partial^4_{x}u   }}
\newcommand{\mycasetwentyTthree}{\ensuremath{ u\partial_{x}u  +   \partial^2_{x}u  +  0.5 \partial^3_{x}u   }}
% =====================================================
\newcommand{\mycasethirtyone}{\ensuremath{ uu_{x} + 1.5 u^2 u_{x} +  u_{xx} + u_{xxxx}  }}
\newcommand{\mycasethirtytwo}{\ensuremath{ uu_{x} + 0.15 u^2 u_{x} +  u_{xx} + u_{xxxx}  }}
\newcommand{\mycasethirtythree}{\ensuremath{ uu_{x} + 0.015 u^2 u_{x} + u_{xx} + u_{xxxx}  }}

\newcommand{\mycasethirtyTone}{\ensuremath{ 0.15 u^2 u_{x} +  u_{xx} + u_{xxxx}  }}
\newcommand{\mycasethirtyTtwo}{\ensuremath{ uu_{x} + 0.15 u^2 u_{x} +  u_{xxxx}  }}
\newcommand{\mycasethirtyTthree}{\ensuremath{ uu_{x} + 0.15 u^2 u_{x} + u_{xx}  }}

% =====================================================
\newcommand{\mycasefourtyone}{\ensuremath{ uu_{x} + u_{xx} + 1.5 u^2 \partial^2_{x}u +  \partial^4_{x}u }}
\newcommand{\mycasefourtytwo}{\ensuremath{ uu_{x} + u_{xx} + 0.15 u^2 \partial^2_{x}u +  \partial^4_{x}u }}
\newcommand{\mycasefourtythree}{\ensuremath{ uu_{x} + u_{xx} + 0.015 u^2 \partial^2_{x}u +  \partial^4_{x}u }}
% =====================================================
\newcommand{\KSc}[3]{\ensuremath{ #1u\partial_{x}u #2\partial^2_{x}u  #3 \partial^4_{x}u  }}

% =====================================================

\newcommand{\caseone}{\ensuremath{ $1$ }}
\newcommand{\casetwo}{\ensuremath{ $2$ }}
\newcommand{\casethree}{\ensuremath{ $3$ }}
\newcommand{\casefour}{\ensuremath{ $-$ }}
\newcommand{\casefive}{\ensuremath{  $-$ }}
\newcommand{\casesix}{\ensuremath{  $-$ }}
\newcommand{\caseseven}{\ensuremath{ $4$ }}
\newcommand{\caseeight}{\ensuremath{ $5$ }}
\newcommand{\casenine}{\ensuremath{ $6$ }}
\newcommand{\caseten}{\ensuremath{  $****$ }}
\newcommand{\caseeleven}{\ensuremath{ $=$  }}
\newcommand{\casetwelve}{\ensuremath{ $7$  }}
\newcommand{\casetwentytwo}{\ensuremath{ $8$  }}
\newcommand{\casetwentyTthree}{\ensuremath{ $9$  }}
\newcommand{\casethirtyTtwo}{{\color{red}\ensuremath{ $10$  }}}
% =====================================================
\newcommand{\ut}{\ensuremath{ \partial_{t}u }}

%\tableofcontents
% ------------------------------
\section{Introduction}
\label{sec:introduction} 

The difference between the solution of the model and the system becomes prominent in many problems involving complex, nonlinear, multi-scale phenomena such as those in engineering~\citep{DeSilva_FAI_2020,Regazzoni_IJNMBE_2021, Wilcox_nature_2021}, thermo-fluids~\citep{Subramanian_AIAA_2020, duraisamy2021perspectives}, and climate/weather prediction~\citep{balaji2021climbing, schneider2021accelerating}; see \citet{Levine_arxiv_2021} for an insightful overview. The deviation of the model from the system is called, in different communities, model uncertainty (structural and/or parametric), model discrepancy, model inadequacy, missing dynamics, or ``\modelerror''; hereafter, the latter is used.

Recently, many studies have focused on leveraging the rapid advances in \gls{ml} techniques and availability of data (e.g., high-quality observations) to develop more accurate models.
Several main approaches include learning fully data-driven (equation-free) models \citep[e.g.,][]{Pathak_PRL_2018, vlachas2018data, dueben2018challenges, weyn2019can, chattopadhyay2020data, arcomano2020machine, chattopadhyay2019analog} or data-driven subgrid-scale closures \citep[e.g.,][]{MA_arxiv_2018,rasp2018deep, maulik2019subgrid2, brenowitz2019spatially, bolton2019applications, beck2019deep, subel2020_sgs, Harlim_JCP_2021, guan2021LES}. In a third approach, corrections to the state or its temporal derivative (tendency) are learned from deviation of the model predictions from the observations \citep{watson2019applying,pawar2020long,pathak2020using,Wattmeyer_GRL_2021,Bretherton_2021_ESSOR}. More specifically, the model is initialized with the observed state, integrated forward in time, and the difference between the predicted state and the observation at the later time is computed. Repeated many times, a correction scheme, e.g., a \gls{dnn}, can be trained to nudge the model's predicted trajectory (or tendency) to that of the system. To deal with observations with noise, a number of studies have integrated \gls{da} with \glspl{dnn}~\citep{Bocquet_NPG_2019, Farchi_RMetS_2020,brajard2020combining, Chen_arxiv_2021,wikner2021using, chattopadhyay2021towards,Gottwald_chaos_2021}. 

These studies often used \glspl{dnn}, showing promising results for a variety of test cases.  However, while powerfully expressive, \glspl{dnn} are currently hard to interpret and often fail to generalize (especially extrapolate) when the systems' parameters change~\citep[e.g.,][]{rasp2018deep,chattopadhyay2020super,guan2021LES, Mojgani_AAAI_2021}.
%This can be alleviated by including interpretable features in the \gls{dnn} architecture~\citep{Mojgani_AAAI_2021}.
The interpretability of models is crucial for robust, reliable, and generalizable decision--critical predictions or designs ~\citep{Wilcox_nature_2021,schneider2021accelerating}. Posing the task of system identification as a linear regression problem, based on a library of nonlinear terms, exchanges the expressivity of \glspl{dnn} for the sake of interpretability~\citep{brunton2016discovering, Rudy_SA_2017}. The closed--form representation of the identified models and their parsimony (i.e., sparsity in the space of the employed library) is the key advantage of these methods, leading to highly interpretable models. A number of studies have used such methods to discover closed--form full models or closures~\citep{Rudy_SA_2017, Zhang_RSP_2018,schaeffer2018extracting,Reinbold_PRE_2020,Zanna_GRL_2020,Messenger_JCP_2021,Cortiella_CMAME_2021}. While the results are promising, \noisy\ data, especially in the chaotic regimes, significantly degrades the quality of the discovered models~\citep{Rudy_SA_2017, Reinbold_PRE_2020,Messenger_JCP_2021,Cortiella_CMAME_2021}.   

So far, there has not been any work on discovering closed--form representation of model error using the differences between model predictions and observations (approach 3) or on {combining} the sparsity-promoting regression methods with \gls{da} to alleviate observation noise. Here, we introduce {\it \method} (Model  Error  Discovery  with  Interpretability  and  Data Assimilation), a general-purpose, data-efficient, non-intrusive framework to discover the structural (and parametric) \modelerrors\ in the form of missing/incorrect terms of \glspl{pde}. \method\ uses differences between predictions from a working numerical solver of the model and noisy sporadic (sparse in time) observations. The framework is built on two main ideas:
\begin{enumerate}
	\item Discovering interpretable and closed--form \modelerrors\ using \gls{rvm}, a sparsity-promoting Bayesian regression method~\citep{Tipping_JMLR_2001},
	\item {Reducing} the effects of {observation} noise using \gls{da} methods such as \gls{enkf} to generate the ``analysis state''.
\end{enumerate}
Below, we present the problem statement and introduce \method. {Subsequently, its} performance is demonstrated using a chaotic \gls{ks} {test case}.  
\section{Problem statement}
\label{sec:problem}
%\input{problem.tex}
%{\color{blue} While \citep{Trehan_IJNME_2017} and $X:2001.05061$ ... }
%In the present \doc, the capitalized and lowercase bold letters denote matrices and vectors respectively, and unbolded letters are used for scalars and functions. 
%The $i^\textit{th}$ column of matrix $\bm{M} \in \realsetT{p}{q}$ is a vector denoted by $\bm{m}_i \in \realsetO{p}$. 
%The $n^\text{th}$ Hadamard power, element-wise power of a matrix, is denoted by $\bm{M}^{\circ n}$, i.e. $\left(\bm{M}^{\circ n}\right)_{i,j} = \bm{m}_{i,j}^n$.
%The Hadamard vector multiplication, element-wise multiplication of two vectors, is denoted by $\bm{m} \odot \bm{n}$, i.e. $\left( \bm{m} \odot \bm{n} \right)_{i,j} = \bm{m}_{i,j} \times \bm{n}_{i,j}$.

Suppose the exact mathematical representation of a system is a nonlinear \gls{pde},
\begin{eqnarray}\label{eq:model_truth}
\partial_t \modelstateparc =  \funtruth{\modelstateparc},
\end{eqnarray}
in a continuous domain. Here, $\modelstateparc$ is the {state variable} at time $t$. While \eqref{eq:model_truth} is unknown, we assume to have access to sporadic pairs of observations of the state variable $u^o$. These observations might be contaminated by measurement noise.
The set of observed states at $\timenow$ is denoted as {$\left\{u^o\left({\ct\timepreti}\right),u^o\left(\timenow\right)\right\}_{i=1}^{\numsamples}$}. Note that $t_i$ do not have to be equally spaced. Furthermore, $\Delta t_i$ should be similar for all $i$ but do not have to be the same (hereafter, we use $\forall i, \Delta t_i=\Delta t$ for convenience).   
%Similar to~\citep{Fablet_IEEE_2018,Bocquet_NPG_2019}, the noise--free observations are considered as the reference and satisfy~\eqref{eq:model_truth}.

Moreover, suppose that we have a \textit{model} of the system,
\begin{eqnarray}\label{eq:model_model}
\partial_t \modelstateparc = \funmodel{\modelstateparc}. %= \funmodel{t,\modelstateparc}  + \funerror{t,\modelstateparc}.
\end{eqnarray}
Without loss of generality~\citep{Levine_arxiv_2021}, we assume that the deviation of~\eqref{eq:model_model} from~\eqref{eq:model_truth} is additive; we further assume that the deviation is only a function of the state~\citep{Bocquet_NPG_2019,Farchi_RMetS_2020}. 
Therefore, the \modelerror\ is
\begin{eqnarray}\label{eq:model_discrepancy}
\funerror{\modelstateparc} := \funtruth{\modelstateparc}- \funmodel{\modelstateparc}.
\end{eqnarray}

Our goal is to find a closed--form representation of $\funerrornull$ given a working numerical solver of \eqref{eq:model_model} and \noisy\ or noise--free observations $\left\{u^o\left(\timepre\right), u^o\left(\timenow\right)\right\}_{i=1}^{\numsamples}$. %Note that with the exception of seeking a closed-form representation, this problem statement is the same as the one addressed in a number of recent studies, which sought an equation-free representation of $h$ \citep[e.g.,][]{watson2019applying, Fablet_IEEE_2018,Bocquet_NPG_2019,Farchi_RMetS_2020}.  

% and similar to~\citep{Farchi_RMetS_2020},

% ------------------------------
\section{Framework: MEDIDA}
\label{sec:method}

\method\ has three main steps (\figref{fig:schematic}): 
Step 1)~Collecting {sporadic} observations of the system and numerically integrating the model forward for short-time (\S\ref{sec:method_error_discovery}); 
Step 2)~Construction and solving a regression problem using \gls{rvm}~\citep{Tipping_JMLR_2001}, which leads to the sparse identification of the \modelerror\ (\S\ref{sec:method_rvm}); Step 3)~If observations are noisy, {following step~1}, stochastic EnKF~\citep{asch2016data,law2015data} is used to estimate an analysis state, an estimate of the system's state, for Step~2 (\S\ref{sec:method_DA}). We emphasize that at no point \method\ needs any knowledge of the system~\eqref{eq:model_truth}. Also, note that other equation-discovery techniques and ensemble-based \gls{da} methods can be used in steps 2-3.   

\subsection{Interpretable \modelerror}
\label{sec:method_error_discovery}
\begin{figure*}[!t]
	\centering
	\begin{minipage}{.45\textwidth}
		\subfloat[Step 1: Observation and model prediction]{
			\includegraphics[trim={0 0 0 0},clip,  scale=0.85]
			{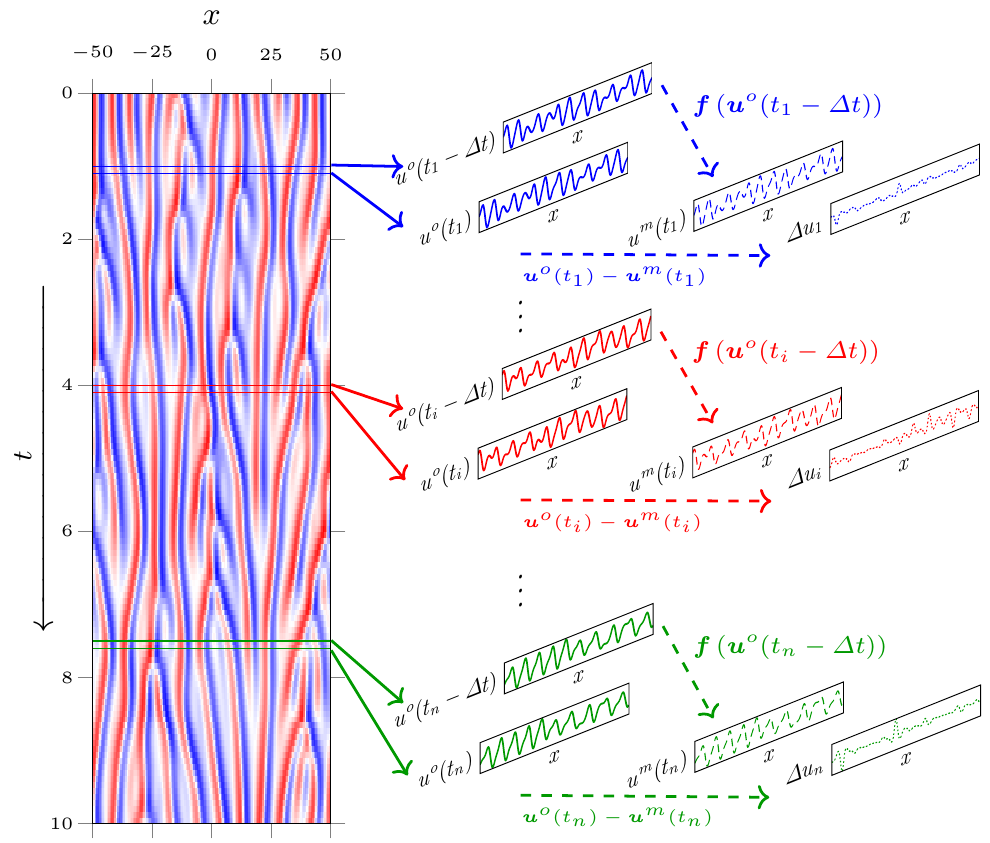}
			\label{fig:discovery}
		}
	\end{minipage}%
	\qquad \qquad
	\begin{minipage}{.45\textwidth}
		\subfloat[Step 2: Forming the regression problem]{
			\includegraphics[trim={0 -1cm 0 -0.4cm},clip, scale=\myscaletwo]
			{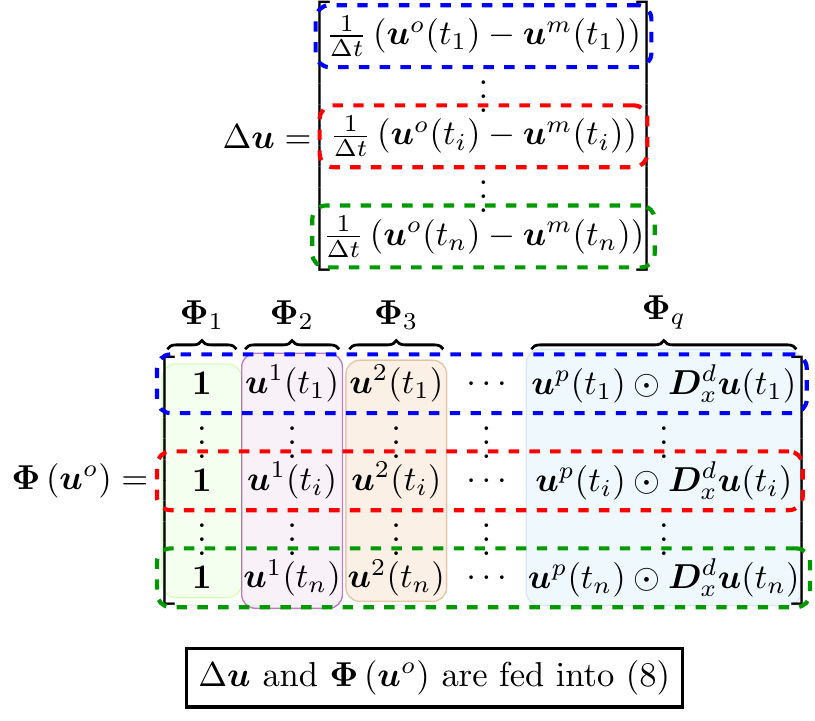}
			\label{fig:discovery2}
		}
	\end{minipage}%
	\newline
	\\
	\begin{minipage}{1.0\textwidth}%	\hspace{-85pt}
		\subfloat[Step 3: \gls{enkf} is used to assimilate \noisy\ observations for step 2]{
			\includegraphics[trim={0 0 4.0cm 0.2cm},clip, width=1.0\textwidth]
			{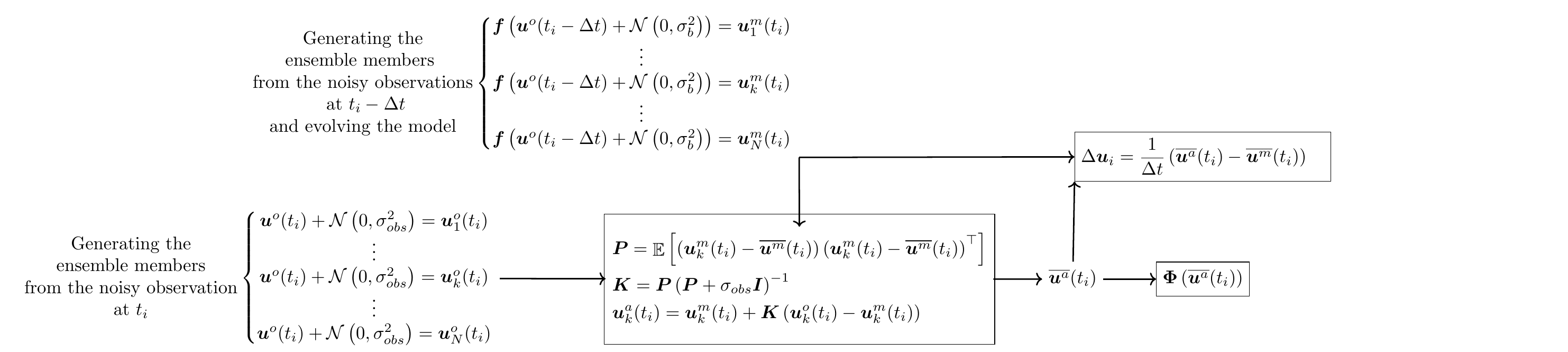}
			\label{fig:assimilation}
		}
	\end{minipage}%
	\caption[]{
		The schematic of MEDIDA, presented in the context of the test case in \S\ref{sec:results}. \subref{fig:discovery} Step~1: $n$ pairs of state variables are sampled uniformly or non-uniformly at $t_i$ from observations (of the system~\eqref{eq:model_truth}) to obtain $\left\{\bm{u}^o\left(\timepre\right), \bm{u}^o\left(\timenow\right)\right\}_{i=1}^{\numsamples}$. The model~\eqref{eq:model_model_dis} is integrated numerically for one time-step from each $\bm{u}^o\left(\timepre\right)$ to predict $\bm{u}^m\left(\timenow\right)$. For each sample, at $t_i$, the difference between the predicted and observed state is used to compute $\Delta \bm{u}_i$ \eqref{eq:diffu_clean}, an approximation of \modelerror. \subref{fig:discovery2} Step~2: $\left\{\Delta \bm{u}_i\right\}_{i=1}^{\numsamples}$ are stacked to form $\Delta \bm{u}$. Moreover, the library of bases $\basesof{\stateobs{}}$, consisting of selected $\numlibcol$ linear and nonlinear bases evaluated at $\left\{\bm{u}^o\left(\timenow\right)\right\}_{i=1}^{\numsamples}$, is formed. Here, the library in \eqref{eq:kernel_dis} is used, but the bases with any arbitrary functional form could be used depending on the anticipated form of the \modelerror. $\Delta \bm{u}$ and $\bm{\Phi}$ are then fed into the \gls{rvm} (or other equation-discovery methods) to compute $\bm{c}^*$ for minimizing the loss function \eqref{eq:model_cost}. The corrected model is then identified using $\bm{c}^*$ as Eq.~\eqref{eq:model_corrected}. \subref{fig:assimilation} Step~3: If observations are noisy, \gls{da} is used to provide an ``analysis state'' $\left\{\bm{u}^a\left(\timenow\right)\right\}_{i=1}^{\numsamples}$ to be used in place of $\left\{\bm{u}^o\left(\timenow\right)\right\}_{i=1}^{\numsamples}$ for computing $\Delta \bm{u}$ and $\bm{\Phi}$ in Steps~1-2. Here we use a stochastic \gls{enkf} for \gls{da}: For each sample $i$, the \noisy\ observation is perturbed by Gaussian white noise with inflated standard deviation $\sigmainflation$ to generate an ensemble of size $\numens$. 
		% Generate ensemble of predicted states
		Each ensemble member $k$ is numerically evolved by the model, 
		$ \left\{  \funmodeldis{\stateobs{k}(\timepre)} \right\}_{k=1}^{\numens}$,
		to generate the ensemble of model prediction 
		$\left\{ \bm{u}_k^m\left( \timenow \right)\right\}_{k=1}^{\numens}$. Observations at time $\timenow$ are also perturbed by Gaussian white noise with standard deviation $\sigmaobs$ to generate the ensemble $\left\{ \bm{u}_k^o \left( \timenow \right) \right\}_{k=1}^{\numens}$. These two ensembles are then used to compute the background covariance matrix $\bm{P}$, Kalman gain $\bm{K}$, and finally, the analysis state $\overline{\bm{u}^a}(t_i)$; overbars indicate ensemble mean and $\mathop{\mathbb{E}} \left[.\right]$ denotes expectation.  $\Delta \bm{u}$ and $\bm{\Phi}$ in~steps 1-2 are then computed using $\overline{\bm{u}^a}(t_i)$.
	}
	\label{fig:schematic}
\end{figure*}

Consider the discretized~\eqref{eq:model_model}:
% ------------------------------------------------------------------
\begin{equation}\label{eq:model_model_dis}
\begin{aligned}
\modelstateparmod{\timenow} = \funmodeldis{\modelstateparobs{\timepre}}.
\end{aligned}
\end{equation}
% ------------------------------------------------------------------
For brevity, we use the notation of an explicit scheme, but the scheme can also be implicit, as shown in the test case. The domain is discretized on the grid of $\grid \in \realsetO{\numgrid}$.  The observation at \timepre, i.e., \modelstateparobs{\timepre}, is the initial condition and 
$\modelstateparmod{\timenow}$ is the predicted state at time $\timenow$ (\figref{fig:discovery}).
%The state of the reference is observed on a discrete grid at each $\timenow$, i.e., $\stateparobs{,\timenow}$.

At each time $\timenow$, subtracting the state predicted by the model ($\modelstateparmod{\timenow}$) from the observed state ($\modelstateparobs{\timenow}$) results in an approximation of the \modelerror\ at $\timenow$ (\figref{fig:discovery2}):
\begin{eqnarray}\label{eq:diffu_clean}
\begin{aligned}
\funerrordis{\timenow}
\approx \Delta \statedis{}_i := 
(\stateobs{}\left(\timenow\right) - \statemodel{}\left(\timenow\right))/\timestep.
\end{aligned}
\end{eqnarray}
{\ct The difference between the states of the system and the model is similar to the analysis increment in the \gls{da} literature, where the best estimate of the state replaces $\stateobs{}\left(\timenow\right)$ in  \eqref{eq:diffu_clean}; see~\S\ref{sec:method_DA}.
	The idea of accounting for model error through the analysis increment was first introduced by \citet{Leith_ARFM_1987}, and is at the core of many recent applications of \gls{ml} aiming to {\it nudge} the model to its correct trajectory \citep{Farchi_RMetS_2020}, or to account for the model error \citep{Carrassi_IJBC_2011,Mitchell_QJRMS_2015}.
	However, in this \doc, we aim to {\it discover} an interpretable representation of model error, $h\left(u\left(t\right)\right)$.
}

Note that to accurately discover {\ct$\funerror{\modelstateparc}$} as defined in \eqref{eq:model_discrepancy}, $\Delta \bm{u}_i$ should be {\it dominated by the \modelerror}, and the contributions from numerical errors (in obtaining $\statemodel{}\left(\timenow\right)$) and measurement errors (in $\modelstateparobs{\timenow}$) should be minimized as much as possible. As discussed in \S \ref{sec:method_DA}, \gls{da} can be utilized to reduce the contributions from the observation noise. Computing $\Delta \bm{u}_i$ after only one $\Delta t$ prevents accumulation of numerical error from integrating \eqref{eq:model_model_dis}. {\ct Moreover, this avoids nonlinear growth of the model error, and its further entanglement with truncation errors, which can complicate the discovery of just model error. Therefore, in \method, the model time step and observation intervals are set to be the same. Note that} while the size of $\Delta t$ could be restricted by availability of the observation pairs, increasing $M$ can be used to reduce truncation errors from spatial {discretizations}.

%Moreover, the model error is approximated by the difference of first--order temporal discretization of the system and model. Consequently, this difference grows nonlinearly in time, and eventually pollutes the discovered terms. In other words, longer evolution of the model inevitably leads to a nonlinear and intractable evolution of model error terms, making approximation of $\funerror{\modelstateparc}$ challenging.}

%Note that by allowing the model and the reference to evolve for a longer simulation/observation time, other sources of error, such as truncation error in the model~\citep{Thaler_JCP_2019}, can dominate the discrepancy between the trajectory of the model and the reference. Therefore, the evolution of the model is limited to a small $\timestep$.

Integrating~\eqref{eq:model_model_dis} from $\modelstateparobs{\timepre}$ and computing $\Delta \statedis{}_i$ from~\eqref{eq:diffu_clean} are repeated for $\numsamples$ samples (step~1), and the vectors of \modelerror\ are stacked to form $\Delta \statedis{} \in \realsetO{\numsamples \numgrid}$ (step~2); see~\figref{fig:schematic}a-b. Similar to past studies~\citep{Rudy_SA_2017,Zhang_RSP_2018,Zhang_JCP_2021,Reinbold_PRE_2020}, we further assume that the \modelerror\ spans over the space of the bases or training vectors, i.e.,
\begin{eqnarray}
%\begin{array}{c}
\funerror{\modelstateparc}=  
%\lincombxc = 
\coeffsymbole_1 \basis_1 + \cdots + \coeffsymbole_\numlibcol \basis_\numlibcol,
%\end{array}%\;,
\label{eq:one}
\end{eqnarray}
where $\basis_i$ is a linear or nonlinear function of the state variable, the building block of the library of training bases
$ %\basesof{\modelstateparc} 
\left\{\basis_k\left({\modelstateparc}\right)\right\}_{k=1}^{\numlibcol}
$. Selection of the bases should be guided by the physical understanding of the system and the model.    

%To form~\eqref{eq:diffu_clean} at each sample $i$, the system is initialized with the observation at $\timepre$ and its evolution is compared to the following observation at $\timenow$ (\figref{fig:discovery}). 
%Subsequently, this procedure is repeated for $\numsamples$ samples and the vectors of \modelerror\ are stacked to form 
% $\Delta \statedis{} \in \realsetO{\numsamples \numgrid}$~(\figref{fig:discovery2}). 
%The samples can be collected sporadically, i.e., at sparse non-uniform temporal intervals.

Here, in the discretized form, we assume that $\basesof{\statepar}$ %, also known as the design matrix,
is a linear combination of polynomial terms up to $\numpoly^\textit{th}$-order and spatial derivatives up to $\numderv^\textit{th}$-order, i.e.,
\begin{eqnarray}
\bases_k \left(\statepar\right) = \statepar^{\powpoly} \odot \bm{D}^\powderv_{x} \statepar; {\powpoly \in \left\{0,\cdots,\numpoly\right\}, \powderv \in \left\{0,\cdots,\numderv\right\}},
\label{eq:kernel_dis}
\end{eqnarray}
where in $\statepar^{\powpoly}$ each element of $\statepar$ is raised to the power of $\powpoly$,
$\bm{D}^\powderv_{x} \statepar$ denotes the $\powderv^\textit{th}$ spatial derivative of $\statepar$, and $\odot $ is the element-wise (Hadamard) product.
%and $\coeff \in \realsetO{\numlibcol}$ is the vector of the coefficients.
Therefore, the \modelerror\ lies on the space of library of the training bases evaluated using the observed state at each $\timenow$, $\basesof{\stateobs{},\timenow} \in \realsetT{\numgrid}{\numlibcol}$. For all the $\numsamples$ samples, the library of the bases are stacked 
to form $\basesof{\stateobs{} } \in \realset{\numsamples \numgrid}{\numlibcol}$~(\figref{fig:discovery2}). 

At the end of step~2, the regression problem for discovery of \modelerror\ is formulated as
% ------------------------------------------------------------------
\begin{equation}\label{eq:model_cost}
\begin{aligned}
\coeffdiscovered =~
& \underset{ \coeff }{arg\minimize}
%& %&  
\NormT{ \Delta \bm{u} - \lincombxuObs}, \\
%& \text{subject to}
%& & c1 \\
%&&& c2
\end{aligned}
\end{equation}
% ------------------------------------------------------------------
where $\coeff \in \realsetO{\numlibcol}$ is the vector of coefficients corresponding to the training bases, and $\coeffdiscovered= \transpose{\left[c^{*}_{1},c^{*}_{2},\cdots,c^{*}_{\numlibcol}\right]}$ is a minimizer of the regression problem, i.e., the vector of coefficients of the \modelerror. Finally, the discovered (corrected) model is identified as
% ------------------------------------------------------------------
\begin{equation}\label{eq:model_corrected}
\begin{aligned}
\partial_t \modelstateparc =
\funmodel{\modelstateparc} + 
\Sigma_{k=1}^{\numlibcol}  c^{*}_{k} \basis_k\left(\modelstateparc\right).
\end{aligned}
\end{equation}
% ------------------------------------------------------------------
%where $\coeffdiscovered = \transpose{\left[c^{*}_{1},c^{*}_{2},\cdots,c^{*}_{\numlibcol}\right]}$.

%\input{schematic.tex}
%\begin{center}
% fig:sample {{{1
% =============================================================================
% }}}1 ========================================================================
%\end{center}

%\newpage
%\twocolumngrid
%\textbf{Remark}: 
%This problem is a variation of the discovery of partial differential equations posed in~\cite{Rudy_SA_2017}, where at the heart of the problem
%\begin{eqnarray}
%%\begin{array}{c}
%\| \bm{u}_t - \lincomb \|^{2}
%%\end{array}
%,
%\label{eq:sindy_cost}
%\end{eqnarray}
%is minimized, given the library of the vectors similar to~\eqref{eq:kernel}.
%
% In~\eqref{eq:sindy_cost}, the tendency of the system, i.e. $\bm{u}_t$, is minimized, where in~\eqref{eq:model_cost} the deviation in state of the system variable is minimized.
%Note that even in~\eqref{eq:sindy_cost}, $\bm{u}_t$ is calculated from approximation of the temporal derivatives based on the descritized observation.\\
%\textbf{Remark}: The trajectory of the imperfect model of~\eqref{eq:model_imperfect} deviates from the perfect model as the system evolve, therefore,  

\subsection{Solution of the regression problem}
\label{sec:method_rvm}
In this study, we use \gls{rvm}~\citep{Tipping_JMLR_2001} to compute $\coeffdiscovered$ in~\eqref{eq:model_cost} for inputs of $\Delta \bm{u}$ and $\basesof{\bm{u}^o}$.
\gls{rvm} leads to a sparse identification of columns of $\bases$ with posterior distribution of the corresponding weights, i.e., the \textit{relevant vectors}. See~\citet{Zhang_RSP_2018} for a detailed discussion in the context of \glspl{pde} discovery.

The breakthrough in equation-discovery originates in introduction of parsimony~\citep{brunton2016discovering}. While the original LASSO-type regularization has yielded promising results in a broad range of applications, \glspl{rvm} have been found to achieve the desired sparsity by a more straightforward hyper-parameter tuning and a relatively lower sensitivity to noise~\citep{Zhang_RSP_2018,Zanna_GRL_2020, Rudy_PhysicaD_2021}.
The hyper-parameter in \glspl{rvm} is a threshold to prune the bases with higher posterior variance, i.e., highly uncertain bases are removed. {To avoid over-fitting, we choose the} hyper-parameter as a compromise between the sparsity and accuracy of the corrected model at the elbow of the L-curve, following~\citet{Mangan_RSA_2017}. %Note that while we have used \gls{rvm} here, other equation-discovery methods can be also used in \method.

%The effect of noise in the sparse identification of nonlinear systems is a topic of ongoing research~\citep{Rudy_SA_2017, Goyal_arxiv_2021,Rudy_PhysicaD_2021,Reinbold_PRE_2020,schaeffer2018extracting}.

\subsection{Data assimilation}
\label{sec:method_DA}
Steps~1-2 described above suffice to find an accurate corrected model \eqref{eq:model_corrected} if {\ct the fully observed state is noise--free}. However, noise can cause substantial challenges in the sparse identification of nonlinear systems, a topic
%that is the subject 
of ongoing research~\citep{Rudy_SA_2017, Goyal_arxiv_2021,Rudy_PhysicaD_2021,Reinbold_PRE_2020,schaeffer2018extracting}.
In \method, we propose to use \gls{da}, a powerful set of tools to deal with \noisy\ {\ct and partial observation (spatio-–temporal sparsity)}. Here, we use stochastic \gls{enkf}~\citep{evensen1994sequential} (\figref{fig:assimilation}). {\ct In this study, we assume that that observations of the full state are available; i.e., $\mathcal{H}(u\left(t\right))= u\left(t\right) \in \realsetO{\numgrid}$ where $\mathcal{H}(.)$ is the observation operator. Dealing with partial observations and more general forms of observation errors (e.g., as in Hamilton et al.~\cite{Hamilton_chaos_2019}) remains subject of future investigations.}
%At each sample $i$, the \gls{enkf} assimilates the \noisy\ observations to the background forecast obtained from the model.
%{For  details of \gls{enkf} see~\cite{SM}.}
%This process for each $\timenow$ is depicted in~\figref{fig:assimilation}.
%Hereafter, the ensemble average of the analysis states acts as the noise--free observation and the ensemble average of the predicted states acts as the model prediction, and are used to construct the regression problem of~\eqref{eq:model_cost}.
%\input{appendix_DA.tex}
{
	%Here, we summarize the \gls{da} method used to infuse the \noisy\ observations to the background forecast.
	
	The result of this step is the analysis state used to construct the vector of model error and the library of the bases (\figref{fig:discovery2}). 
	
	At each sample time $\timepre$, the observations are further perturbed with Gaussian white noise to obtain an ensemble of size $\numens$ of the initial conditions,
	\begin{eqnarray}\label{eq:DA_ensemble_initial}
	\stateobs{k}(\timepre) =  \stateobs{}(\timepre)  +  \disNormal{0}{\sigmainflation^2},
	\end{eqnarray}
	where $\sigmainflation$ is  standard deviation of the observation noise ($\sigmaobs$) times an inflation factor \citep{anderson1999monte,asch2016data}, and $k$ denotes the $k^{\textit{th}}$ ensemble member.
	
	Subsequently, the model is evolved for each of these ensemble members, i.e., ${\bm{u}_k^m}(\timenow)=\funmodeldis{\stateobs{k}(\timepre)}$.
	This ensemble of model prediction is used to construct the background covariance matrix as% }
	\begin{eqnarray}
	\label{eq:EnKF_P}
	\begin{aligned}
	\bm{P}&= \mathop{\mathbb{E}}\left[
	\left(\bm{u}_k^m(\timenow)-\overline{\bm{u}^m}(\timenow)\right)
	\transposeParan{\bm{u}_k^m(\timenow)-\overline{\bm{u}^m}(\timenow)}
	\right], \\
	\end{aligned}
	\end{eqnarray}
	where $\overline{\bm{u}^m}(\timenow)$ is the ensemble average of  the model prediction, and 
	$\mathop{\mathbb{E}} \left[.\right]$ denotes the expected value.
	Having the observation operator to be linear, and non-sparse,
	the Kalman gain,  $\bm{K} \in \realsetT{\numgrid}{\numgrid}$, is then calculated as,
	\begin{eqnarray}
	\label{eq:EnKF_E}
	\begin{aligned}
	\bm{K}&=\bm{P}  \inv{ \left( \bm{P} + \sigmaobs\bm{I} \right) },\\
	\end{aligned}
	\end{eqnarray}
	where $\bm{I} \in \realsetT{\numgrid}{\numgrid}$ is the identity matrix.
	
	For each sample at the prediction time $\timenow$, the $k^\textit{th}$ member of the ensemble of the observations is generated as
	\begin{eqnarray}\label{eq:DA_ensemble_analysis}
	\stateobs{k}(\timenow) =  \stateobs{}(\timenow)  +  \disNormal{0}{\sigmaobs^2}.
	\end{eqnarray}
	{\ct
		Note that while the \gls{enkf} can be used for non--Gaussian observation noise, the method would be sub--optimal and may require further modification~\citep{Carrassi_WIRES_2018}. Hereon, we limit our experiments to Gaussian noise.}
	
	Finally, the $k^\textit{th}$ member of the ensemble of the analysis state is 
	\begin{eqnarray}
	\label{eq:EnKF}
	\begin{aligned}
	\stateanalysis{k}(\timenow) =
	\statemodel{k}(\timenow)+\bm{K}\left(\stateobs{k}(\timenow) - \statemodel{k}(\timenow)  \right).
	\end{aligned}
	\end{eqnarray}
	The process of \gls{enkf} assimilates the \noisy\ observations to the background forecast obtained from the model, $\statemodel{k}(\timenow)$.
	Subsequently, the ensemble average of the analysis states, $\overline{\stateanalysis{}}(\timenow)$, is treated as {\ct a close estimate of the true} state of the system, $\stateobs{}\left(\timenow\right)$ in \S\ref{sec:method_error_discovery}.
	
	Similarly, the difference between the ensemble averages of  the analysis state and the model prediction, i.e.,
	\begin{eqnarray}
	\label{eq:diffu_DA}
	\Delta \bm{u}(\timenow) =(
	\overline{\stateanalysis{}}(\timenow)-\overline{\statemodel{}}(\timenow))/\Delta t,
	\end{eqnarray}
	replaces~\eqref{eq:diffu_clean} and is fed into the algorithm of \S\ref{sec:method_error_discovery}.
	
	{\ct There are a few points about this \gls{da} step that need to be further clarified. First, as discussed in \S\ref{sec:method_error_discovery}, the model is evolved for only one time step. Although this may hinder the traditional spin--up period used in \gls{enkf}, as a compromise, we resort to a large ensemble and inflation to ensure successful estimation of the state. 
		
		Second, for MEDIDA to work well, \eqref{eq:diffu_DA} should be dominated by model error. However, the inevitable presence of other sources of error, notably the observation errors, generally leads to an overestimation of the actual model error; see~\citet{Carrassi_IJBC_2011} for discussions and remedies.}
	
	Finally, note that \gls{enks}~\citep{Evensen_MWR_2000} is shown to be more suitable for offline pre-processing of the data used in \gls{ml} models~\citep{Chen_arxiv_2021};
	however; in \method, the model is evolved for a single time step, {\ct therefore, we expect that \gls{enks} and \gls{enkf} to behave similarly.}}

\subsection{Performance metric}
\label{sec:method_error}

To quantify the accuracy of \method, we use the normalized distance between the vector of coefficients, {\ct defined as}
\begin{eqnarray}
	\errorcoeffmodel = \frac{ \NormT{ \coefftrue - \coeffmodel  } }{ \NormT{ \coefftrue } }
	\; \; \mathrm{and} \; \;
	\errorcoeffdiscovered = \frac{ \NormT{ \coefftrue - \coeffdiscovered  } }{ \NormT{ \coefftrue } },
	\label{eq:error_coeff}
\end{eqnarray}
{\ct where $\coefftrue$, $\coeffmodel$, $\coeffdiscovered$ are coefficient vectors of the system \eqref{eq:model_truth}, the model \eqref{eq:model_model}, and the corrected model \eqref{eq:model_corrected}\citep{Rudy_SA_2017, Reinbold_PRE_2020}. The $i^\textit{th}$ element of the coefficient vector is a scalar corresponding to the $i^\textit{th}$ term in the equation.} Quantitatively, the goal is to achieve $\errorcoeffdiscovered \ll \errorcoeffmodel$.
{\ct Note that the implicit assumption in this metric is that the model and the model error can be expressed on the space spanned by the bases of the library.}
% ------------------------------
\section{Test case: KS equation}
\label{sec:results}
%\input{results.tex}
%\subsection{Kuramoto–Sivashinsky equation}
To evaluate the performance of \method, we use a chaotic \gls{ks} system, a challenging test case for system identification, particularly from noisy observations~\citep{Rudy_SA_2017,Reinbold_PRE_2020}. The \gls{pde} is 
\begin{eqnarray}
	\ut \KSc{+}{+}{+}=0,
	\label{eq:KS}
\end{eqnarray}
where $u\partial_{x}u$ is convection, $\partial_{x}^2{u}$ is anti-diffusion (destabilizing), and $\partial_{x}^4{u}$ is hyper-diffussion (stabilizing). 
%This combination leads to small-scale dissipations and large-scale instabilities, transferring energy from the large to small scales~\citep{Edson_ANZIAM_2019, Sprott_2010}. 
The domain is periodic, $u\left(0,t\right)=u\left(L,t\right)$. We use $L=32\pi$, which leads to a highly chaotic system~\citep{Pathak_PRL_2018,khodkar2021data}. %in~\figref{fig:discovery}.  
%Fore more details regarding the solution of~\eqref{eq:KS} see~\cite{SM}.
%The KS equation demonstrates a challenging case of chaotic behavior, especially for a system identification task and in the presence of noise~\citep{Rudy_SA_2017,Reinbold_PRE_2020}.

Here, \eqref{eq:KS} is the system. We have created 9 (imperfect) models as shown in Table~\ref{tab:cases}. In cases \caseone-\casethree, one of the convection, anti-diffusion, or hyper-diffusion term is entirely missing (i.e., structural uncertainty). In cases \caseseven-\casenine, some or all of the coefficients of the system are incorrect (parametric uncertainty). Finally, in cases \casetwelve-\casetwentyTthree, a mixture of parametric and structural uncertainties, with missing and extra terms, is present. 

The system and the models are numerically solved  using different time-integration schemes and time-step sizes to introduce additional challenges and a more realistic test case. To generate $\bm{u}^o$, \eqref{eq:KS} is integrated with the exponential time-differencing fourth-order Runge-Kutta~\citep{Kassam_SIAM_2005} with time-step $\Delta t^o$ (in \figref{fig:discovery}, $1t=10^3\Delta t^o$). The models are integrated with second-order Crank-Nicolson and Adams-Bashforth schemes with $\Delta t^m=\Delta t=5 \Delta t^o$. For both system and models, $\numgrid=1024$ Fourier modes are used  (different discretizations or $\numgrid$ for system and models could have been used too). $\bm{u}^o$ is collected after a spin-up period of $\tau=200\Delta t$ and uniformly at $t_i=i \tau$. $\bases$, is constructed with $\numderv=4$ and $\numpoly=4$ as defined in~\eqref{eq:kernel_dis}. Open-source codes are used to generate $\bases$~\citep{Rudy_SA_2017} and solve \eqref{eq:model_cost} using \gls{rvm}~\citep{Zanna_GRL_2020}. 

%In both the reference and the models, $\numgrid=1024$ Fourier modes are used. % In practice, the frequency of data measurements ($\timestep^o$) does not necessarily match the time steps used in the model ($\timestep^m$).
%\textbf{Remark}: In CNAB2 the numerical simulation of KS, the linear and nonlinear terms are descritized differently compared to~\eqref{eq:model_cost}.
%The possibility of the error discovery where the numerical discretization does not match the discretization scheme of the cost function showcases the robustness of the proposed framework.
%The reference and the models are evolved with $\timestep^o = 10^{-4}$ and  $\timestep^m = 5\timestep^o$, respectively.

%% fig:KS {{{1
%% =============================================================================
%\begin{figure}[!htbp]
%	\centering
%	\includegraphics[scale=1.5]{data/KS.pdf}
%	\caption{Solution of KS}
%	\label{fig:KS}
%\end{figure} 
%% }}}1 ========================================================================

\subsection{Noise--free observations}
\label{sec:KS_clean}
First, we examine the performance of \method\ for noise--free observations (only steps 1-2). The motivation is two--fold: 
i) to test steps 1-2 before adding the complexity of noisy observations/\gls{da}, and 
ii) in many applications, an estimate of $\bm{u}^o$ is already available (see \S\ref{sec:discussion}). 
%For example, a high-fidelity (but computationally very expensive) model can be treated as a surrogate of the system and its (noise--free) simulation outputs can be be used as $\bm{u}^o$. As another example, sometimes the analysis state is routinely generated and would be available to be used as $\bm{u}^o$; this is often the case for atmosphere and some other components of the Earth system. Both examples are highly relevant to climate/weather modeling, and used in some recent studies using \glspl{dnn} for model error correction~\citep[e.g.,][]{Wattmeyer_ESSOr_2021, Bretherton_2021_ESSOR}.

Here, $\bm{u}^o$ are from the numerical solutions of~\eqref{eq:KS}. 
The models, \method-corrected models, and the corresponding errors for $\numsamples = 100$ samples show $\errorcoeffdiscovered<2\%$, between 40 to 200 times improvement compared to $\errorcoeffmodel$ (Table~\ref{tab:cases}).

{To conclude, }\method\ is a data-efficient framework. Its performance is insensitive to $\numsamples$, such that $\errorcoeffdiscovered$ changes by $0.1\%$ for $\numsamples \in \left\{10,100,1000\right\}$. There are two reasons for this: i) \gls{rvm} is known to be data-efficient (compared to \glspl{dnn}) \citep{bishop2006pattern}, and ii) each grid point at $\timenow$ provides a data-point, i.e., a row in $\Delta \bm{u}$ and $\bm{\Phi}$, although these are not all independent/uncorrelated data-points.

%In our experiments, the corrected models of all the cases converge to the truth as $\timestep^m \rightarrow \timestep^o$.

%\begin{center}
% fig:discoery: clean and noisy {{{1
% =============================================================================
\begin{table*}[!t]
	\caption{\method-corrected models from noise--free and noisy observations of the KS system: $0=\ut\KSc{+}{+}{+}$. $\errorcoeffmodel$ and $\errorcoeffdiscovered$ are defined in \eqref{eq:error_coeff}.
		For the \noisy\ cases, $\sigmainflation=20\sigma_{obs}$, and the ensemble size is $\numens=10\numgrid$, except for cases $\casethree$, $\casenine$, and $\casetwentyTthree$ with $\numens=200\numgrid$, $100\numgrid$, and $500\numgrid$, respectively.
	}
	\label{tab:cases}
	\centering
	\begin{tabular}{ |p{0.5cm}|
			%p{1.0cm}
			p{4.0cm}|p{0.90cm}|p{4.5cm}|p{0.90cm}|p{4.5cm}|p{0.90cm}| }
		%\hline
		%\multicolumn{7}{|c|}{System: $ \ut\KSc{+}{+}{+}=0$ } \\  
		\hline
		\# 
		%& Sample size 
		& Model: \hspace{90pt}$0 = \ut + $
		& $\errorcoeffmodel \left[\%\right]$
		& Corrected model (noise--free): \hspace{20pt} $0 = \ut +$
		%& Error bar
		& $\errorcoeffdiscovered \left[\%\right]$
		& Corrected model (noisy): \hspace{30pt} $0 = \ut +$
		%& Error bar
		& $\errorcoeffdiscovered \left[\%\right]$
		\\
		\hline
		\caseone %& $1000$
		& \mycaseone
		& $57.74$
		& \KSc{\numprint{0.9740}}{+}{+}%$0.9740 uu_x{} + u_{xx} + u_{xxxx} $ 
		%& $1.15\times10^{-9}$
		& $1.50$
		& $\KSc{0.98}{+}{+}$ 
		& $1.43$
		\\
		%\hline
		\casetwo  %& $1000$
		& \mycasetwo
		& $57.74$
		& \KSc{}{+\numprint{0.9933}}{+}%$ uu_{x} + 0.9933 u_{xx} + u_{xxxx}$ 
		%& $1.79\times10^{-10}$
		& $0.39$
		& $\KSc{}{+0.98}{+}$ 
		& $0.98$
		\\
		%\hline
		\casethree  %& $1000$
		& \mycasethree
		& $57.74$
		& \KSc{}{+}{+\numprint{1.0187}}%$uu_{x} + u_{xx} + 1.0187 u_{xxxx}$ 
		%& $1.93\times10^{-10}$
		& $1.08$
		& $\KSc{}{+}{+1.02}$ 
		& $0.98$
		\\
		%\hline
		\caseseven %& $1000$
		& \mycaseseven 
		& $28.87$
		& \KSc{\numprint{0.9870}}{+}{+}%$0.9870 uu_{x} + u_{xx} + u_{xxxx}	$ 
		%& $1.15\times10^{-9}$
		& $0.75$
		& $\KSc{0.99}{+}{+}$ 
		& $0.63$
		\\
		%\hline
		\caseeight  %& $1000$
		& \mycaseeight 
		& $28.87$
		& \KSc{}{+\numprint{0.9976}}{+}%$uu_{x} + 0.9976 u_{xx} + u_{xxxx}$ 
		%& $9.76\times10^{-11}$
		& $0.14$
		& $\KSc{}{+0.98}{+}$ 
		& $0.99$	
		\\
		%\hline
		\casenine  %& $1000$
		& \mycasenine 
		& $28.87$
		& \KSc{}{+}{+\numprint{1.0025}}%$uu_{x} + u_{xx} + 1.0025 u_{xxxx}5$ 
		%& $1.42\times10^{-10}$
		& $0.14$
		& $\KSc{}{+}{+1.00}$ %"$+1.0uu_{x}+1.0u_{xx}+1.0033u_{xxxx}$"
		& $0.15$
		\\
		%\hline
		\casetwelve  %& $1000$
		& \mycasetwelve
		& $86.60$
		& \KSc{\numprint{0.9732}}{+\numprint{0.9995}}{+}%$	0.9732uu_{x}+0.9995u_{xx}+u_{xxxx}	$
		%& $.\times10^{-}$
		& $1.55$
		& $\KSc{0.98}{+0.98}{+}$ 
		& $1.63$
		\\
		%\hline
		\casetwentytwo  %& $1000$
		& \mycasetwentytwo
		& $28.87$
		& \KSc{}{+}{+}%+0.0024u_{xxx}
		%& $.\times10^{-}$
		& $0.28$
		& $u\partial_{x}u  +  \partial^2_{x}u - \numprint{0.0144} \partial^3_{x}u + \partial^4_{x}u$
		% $+1.0uu_{x}+1.0u_{xx}+0.0144u_{xxx}+1.0u_{xxxx}$
		& $0.83$
		\\
		%\hline
		\casetwentyTthree  %& $1000$
		& \mycasetwentyTthree
		& $64.55$
		&  $u\partial_{x}u  +  \partial^2_{x}u -\numprint{0.0131} \partial^3_{x}u + \numprint{1.0168}\partial^4_{x}u$
		& $1.29$
		& $u\partial_{x}u  +  \partial^2_{x}u - \numprint{0.0095} \partial^3_{x}u + \numprint{1.0014}\partial^4_{x}u$ %$+1.0uu_{x}+1.0u_{xx}-0.0095u_{xxx}+1.0014u_{xxxx}$
		& $0.55$
		\\
		\hline
	\end{tabular}
\end{table*}
% }}}1 ========================================================================
%\end{center}
%}

\subsection{\Noisy\ observations}
\label{sec:KS_noisy}
\begin{figure}[!t]
	\centering
	\includegraphics[scale=\myscalethree]{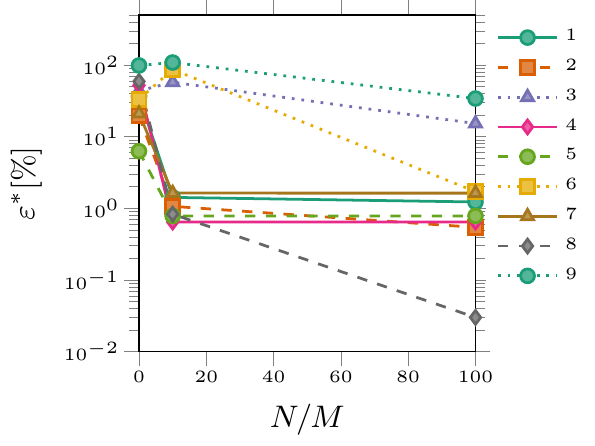}
	\caption{
		Improvement in the performance of \method\ as the ensemble size, $\numens$, is increased ($n=100$). $N/M=0$ corresponds to no \gls{da}. $\errorcoeffdiscovered$ for cases $\casethree$ and $\casetwentyTthree$ further drops to $0.98\%$ and $0.55\%$ using larger ensembles of $\numens=200\numgrid$ and  $\numens=500\numgrid$, respectively (Table~\ref{tab:cases}).
	} 
	\label{fig:RVM_noise_1d0}
\end{figure}

Next, we examine the performance of \method\ for \noisy\ observations, obtained from adding noise to the numerical solution of \gls{ks}: $
%\begin{eqnarray}\label{eq:additive_noise}
\modelstateparobsc =  \modelstateparc  +  \disNormal{0}{\sigmaobs^2}.
%\end{eqnarray}
$
Here, $\noiseratio = 0.01$, where $\sigma_{u}$ is the standard deviation of the state {(equivalent to $0.1$ following the methodology in~\citep{Reinbold_PRE_2020,Kaheman_arxiv_2020})}. Without step~3 (\gls{da}), the \modelerror\ discovery fails, leading to many spurious terms and $\errorcoeffdiscovered$ of $O(10\%)-O(100\%)$, comparable or even worse than {the model error,} $\errorcoeffmodel$ (\figref{fig:RVM_noise_1d0}). 

Table~\ref{tab:cases} shows the corrected models (with step~1-3). With $\numens=10\numgrid$, in all cases except for 3, 6, and 9, $\errorcoeffdiscovered < 2\%$,  between 30 to 60 times lower than $\errorcoeffmodel$. For those 3 cases, increasing $N$ further reduces $\errorcoeffdiscovered$ (\figref{fig:RVM_noise_1d0}), leading to the discovery of accurate models with $\errorcoeffdiscovered < 1\%$ at large ensembles with $\numens=100\numgrid-500\numgrid$~(Table~\ref{tab:cases}). Note that one common aspect of these cases is that the model is missing or mis-representing the hyper-diffusion term. The larger $\numens$ is needed to further reduce the noise in the analysis state to prevent amplification of the noise due to the higher-order derivative of this term. It should also be highlighted that while $\numens$ at the order of $\numgrid$ or larger might seem impractical, each ensemble member requires only {\it one} time-step integration (thus, a total of $\numsamples\numens$ time steps).

%Note that in \gls{enkf}, $\bm{u}_k^o(\timepre)$ are evolved using the model. Here, models have large structural/parametric errors to demonstrate the power of \method; however, this can lead to slow convergence of \gls{enkf}. In all cases examined here, the convergence could be achieved by increasing $N$ (Fig.~\ref{fig:RVM_noise_1d0}). 

\method\ is found to be data-efficient in these experiments too. Like before, $\errorcoeffdiscovered$ changes by $0.1\%$ for $\numsamples \in \left\{10,100,1000\right\}$ for all cases except for 3, 6, and 9. For these 3 cases, $\errorcoeffdiscovered$ improves by about $10\%$ as $\numsamples$ is increased from $10$ to $100$, but then changes by only $0.1\%$ when $\numsamples$ is further increased to $1000$.

%However, even in these more challenging with slow convergence, increasing the ensemble size leads the discovery of \modelerror~(Table~\ref{tab:noisy_1d0_allcases}).

%These cases are also sensitive to the number of samples, and results are shown for $\numsamples$ = 100.
%Note that generation of ensemble of size $\numens$ requires only one time step evolution of the model. 

%The models are corrected from observation with $\noiseratio = 0.01$ using ensemble size of $\numens=10\numgrid$, in~Table~\ref{tab:noisy_1d0_allcases} (for $\noiseratio = 0.005$ see \citep{SM}).

%In~\figref{fig:RVM_noise_1d0}, the convergence of \method\ with respect to the ensemble size is shown. 
%Consider the \modelerror\ at  $\numens=0$ as for the baseline. 

%As the result of a trial and error, the inflation factor is set to $\sigmainflation=0.25$.

%For all models except \casethree, \casenine, and \casetwentyTthree, the error reach to the $2\%$ error threshold with  $\numens=10\numgrid$. 

% fig:noise {{{1
% =============================================================================
% }}}1 ========================================================================

\section{Discussion and Summary}
\label{sec:discussion}
%\input{discussion.tex}
%We discuss the major findings of identification of \modelerror.
We introduced \method, a data--efficient, and non--intrusive framework to discover interpretable, closed--form structural (or parametric) \modelerror\ for chaotic systems. \method\ only needs a working numerical solver of the model and $\numsamples$ sporadic pairs of noise--free or noisy observations of the system. \Modelerror\ is estimated from differences between the observed and predicted states (obtained from one--time--step integration of the model from earlier observations). A closed--form representation of the \modelerror\ is estimated using an equation-discovery technique such as \gls{rvm}. \method\ is expected to work accurately if i) the aforementioned differences are dominated by \modelerror\ and not by numerical ({discretization}) error or observation noise, and ii) the library of the bases is adequate (further discussed below). Regarding (i), the numerical error can be reduced by using higher numerical resolutions while the observation noise can be tackled using \gls{da} techniques such as \gls{enkf}. 

The performance of \method\ is demonstrated using a chaotic \gls{ks} system for both noise--free and noisy observations. In the absence of noise, even with a small $\numsamples$, \method\ shows excellent performance and accurately discovers linear and nonlinear model errors, leading to corrected models that are very close to the system. These results already show the potential of \method\ as in some important applications, noise--free estimates of the system ({\ct re}analysis states) are routinely generated and are available; this is often the case for atmosphere and some other components of the Earth system{\ct \cite{ERALAND_2021, ERA5_2020, Fan_JoC_2006}}. For example, such {\ct re}analysis states have been recently {\ct used to correct the \modelerrors\ of an \gls{nwp} model \citep{Wattmeyer_GRL_2021},
	a \gls{qg} model
	\citep{Farchi_RMetS_2020}, and a \gls{mooam} \citep{brajard2020combiningSGS}}.

% On library:

%Pruning the library based on our \apriori\ knowledge of the system, or expanding it are beyond the scope of the present \doc.

% On stability:

%In the absence of noise, the closed--form, interpretable models of the error are discovered from a limited number %sporadically observed samples.
%To pose the regression problem, the observations of the system are collected and the model is evolved from one %time level to the following time level.
%By solving the regression problem, using relevance vector machine, the missing terms are identified. 
%This results are specifically important since,\mynobreakpar
%\begin{enumerate}
%\item the method does not require numerous samples, often necessary to train a neural network based model %nudging/correcting terms,
%\item the correction has an interpretable closed--form representing the \modelerror, and
%\item the identified terms represent a missing or inaccurately modeled physics.
%, either due to modeling or lack of knowledge of the system
%\end{enumerate}

%The minor difference between the reference and the model, can be associated to the truncation error, which are of higher order derivatives than of those present in the library of the bases. 
%The discovery of the truncation error is possible using the same approach and only by extending the library~\citep{Thaler_JCP_2019}, in that case the truncation terms could be distinguished from structural error by analysis of modified differential equations~(MDEs).

In the presence of noise, the \modelerror\ could not be accurately discovered without \gls{da}. Once \gls{enkf} is employed, \method\ accurately discovers linear and nonlinear model errors. A few cases with model errors involving linear but $4^\textit{th}$-order derivatives require larger ensembles. This is because higher-quality analysis states are needed to avoid amplification of any remaining noise as a result of high-order derivatives.
%As mentioned in~\S\ref{sec:KS_noisy}, producing a large ensemble is not necessarily computationally demanding here, since each member only requires one time-step integration.

{\ct
	Although the number of ensemble members used in \method\ is more than what it is commonly used in operational DA, the associated cost is tractable, since each ensemble is evolved only for one time step.  This is another advantage of evolving the model for only one time step, which as discussed before, and is also motivated by reducing the accumulation of truncation errors and nonlinear growth of the model errors.}
%  This limitation does not exist in corrections based on \glspl{dnn}, where the networks are trained to account for both the model error and the nonlinear evolution of the error\citep{Carrassi_IJBC_2011,Mitchell_QJRMS_2015,Farchi_RMetS_2020}.}
Still, if needed, inexpensive surrogates of the model that could provide accurate one--time--step forecasts can be used to efficiently produce the ensembles~\citep{chattopadhyay2021towards,Pathak_arxiv_2022}.

Also, note that in \gls{enkf}, ensemble members are evolved using the model. If the \modelerror\ is large, the analysis state might be too far from the system's state, degrading the performance of \method. In all cases examined here, even though the \modelerrors\ are large, with large-enough ensembles, the {approximated} analysis states are accurate enough to enable \method\ to {discover the missing or inaccurate terms}. However, in more complex systems, this might become a challenge that requires devising new remedies. One potential resolution is an iterative procedure, in which the analysis state is generated and a corrected model is identified, which is then used to produce a new analysis state and a new corrected model, and this continues until convergence. {Although such iterative approaches currently lack proof of convergence, they have shown promises in similar settings~\citep{brunton2016discovering,Schaeffer_RSA_2017,Hamilton_chaos_2019, Schneider_arxiv_2020}; for example,~\citet{Hamilton_chaos_2019} update the error in the observations iteratively until the convergence of the estimated state from the \gls{enkf}}.
The corresponding cost and possible convergence {properties} of such an iterative approach for \method remain to be investigated. {Similarly,} other methods for dealing with noise in equation discovery~\citep[e.g.,][]{Reinbold_PRE_2020,Reinbold_NComm_2021,Goyal_arxiv_2021} could also be integrated into \method.

%The challenge in these cases is the different rate of convergence of differences bases in the library, and to calculate an accurate high-order derivative of the analysis state from the \noisy\ observations. In such cases increasing the size of the ensemble is necessary to correctly identify the \modelerror. Reformulating the problem in its weak form~\citep{Reinbold_PRE_2020,Reinbold_NComm_2021} can be one approach to further alleviate the sensitivity to noise, however, it is restricted to the discovery of linear terms~\citep{Reinbold_PRE_2020}.

% On library
The choice of an adequate library for \gls{rvm} is crucial in \method. Although an exhaustive library of the training vectors with any arbitrary bases is straightforward, it quickly becomes computationally intractable. Any \apriori\ knowledge of the system, such as locality, homogeneity~\citep{Bocquet_NPG_2019}, Galilean invariance, and conservation properties~\citep{Zhang_RSP_2018} can be considered to construct a concise library. Conversely, the library can be expanded to ``explore the computational universe'', e.g., using gene expression programming~\citep{Vaddireddy_PF_2020}. Even further, additional {constraints, such as} stability of the corrected model, can be imposed~\citep{Mojgani_IJNME_2020,Loiseau_JFM_2018,Chen_JCP_2020}. Effective strategies for the selection of an adequate and concise library can be investigated in future work using more complex test cases.

% On masked data
{Finally, beyond dealing with noisy observations, another challenge in the discovery of interpretable model errors is approximating the library given spatially sparse observations. Approaches that leverage auto-differentiation for building the library in equation discovery have recently shown promising results~\citep{Chen_NatC_2021}, and can be integrated into \method. 
	
	{\method} is shown to work well for a widely used but simple chaotic prototype, the \gls{ks} system. The next step will be investigating the performance of {\method} for more complex test cases, such as a two-layer quasi-geostrophic model. Scaling {\method} to more complex, higher dimensional systems will enable us to discover interpretable model errors for \glspl{gcm}, \gls{nwp} models, and other models of the climate system.    
}

\begin{acknowledgments}
We are grateful to Yifei Guan for helpful comments on the manuscript. {\ct We also would like to thank two anonymous reviewers for their constructive and insightful comments.} This work was supported by an award from the ONR Young Investigator Program (N00014-20-1-2722), a grant from the NSF CSSI program (OAC-2005123), and by the generosity of Eric and Wendy Schmidt by recommendation of the Schmidt Futures program. Computational resources were provided by NSF XSEDE (allocation ATM170020) and NCAR’s CISL (allocation URIC0004). Our codes and data are available at~\mydepository.
\end{acknowledgments}
%\acknowledgments{
%We thank Yifei Guan for helpful comments on the manuscript. This work was supported by an award from the ONR Young Investigator Program (N00014-20-1-2722), a grant from the NSF CSSI program (OAC-2005123), and by the generosity of Eric and Wendy Schmidt by recommendation of the Schmidt Futures program. Computational resources were provided by NSF XSEDE (allocation ATM170020) and NCAR’s CISL (allocation URIC0004). Our codes and data are available at~\mydepository.
%}

% New version of the num-names style
\bibliographystyle{elsarticle-num-names}
%\bibliography{ModelError}{}
%\nocite{*}
\bibliography{ModelError}% Produces the bibliography via BibTeX.

\newpage

%\appendix*
\setcounter{table}{0}
\renewcommand{\thetable}{S.\Roman{table}}%
\setcounter{figure}{0}
\renewcommand{\thefigure}{S.\Roman{figure}}%
\setcounter{equation}{0}
\renewcommand{\theequation}{S.\arabic{equation}}%
%\glsresetall

%\section*{Supplemental Material}

%\subsection{Relevance vector machines (RVMs)}
%\label{sec:appendix_method_rvm}
%\input{appendix_RVM.tex}

%\subsection*{\Gls{enkf}}
%\label{sec:appendix_method_DA}
%\input{appendix_DA.tex}

%\subsection{Solution of KS}
%\label{sec:appendix_KS}
%\input{appendix_KS.tex}

%\subsection{Additional experiments}
%\label{sec:appendix_results}
%\input{appendix_results.tex}

\end{document}